\documentclass[12pt]{article}
\pdfoutput=1
\usepackage{putex}
\usepackage{graphicx}
\usepackage{epstopdf}
\usepackage{enumerate}
\usepackage{cite}
\usepackage{tensor}
\usepackage{slashed}

\numberwithin{equation}{section}

\newcommand{\abs}[1]{\left\lvert #1 \right\rvert}

\newcommand {\be} {\begin {equation}}
\newcommand {\ee} {\end {equation}}

\newcommand {\bes} {\begin {equation*}}
\newcommand {\ees} {\end {equation*}}

\newcommand{\es}[2] {\begin{equation} \label{#1} \begin{split} #2 \end{split} \end{equation}}

\newcommand{\CP}{\mathbb{CP}}
\newcommand{\Z}{\mathbb{Z}}

\newcommand{\R}{\mathbb{R}}

\newcommand{\myn}{n}

\newcommand{\beq}{\begin{equation}}
\newcommand{\eeq}{\end{equation}}

\begin{document}

\preprint{MIT-CTP-4450}

\institution{MIT}{Center for Theoretical Physics, Massachusetts Institute of Technology, Cambridge, MA 02139}
\institution{Princeton}{Joseph Henry Laboratories, Princeton University, Princeton, NJ 08544}

\title{Anomalous dimensions of monopole operators in three-dimensional quantum electrodynamics}

\authors{Silviu S.~Pufu}

\abstract{
The space of local operators in three-dimensional quantum electrodynamics contains monopole operators that create $n$ units of gauge flux emanating from the insertion point.  This paper uses the state-operator correspondence to calculate the anomalous dimensions of these monopole operators perturbatively to next-to-leading order in the $1/N_f$ expansion, thus improving on the existing leading order results in the literature.  Here, $N_f$ is the number of two-component complex fermion flavors.  The scaling dimension of the $n=1$ monopole operator is $0.265 N_f - 0.0383 + O(1/N_f)$ at the infrared conformal fixed point.}
\date{Januray, 2014}

\maketitle

%\tableofcontents

\section{Introduction}

There are several reasons why one may be interested in studying quantum electrodynamics in three dimensions (QED$_3$).  To a high energy physicist, QED$_3$ may be of interest because it bears some resemblance to four-dimensional quantum chromodynamics (QCD$_4$).  Indeed, both theories are asymptotically free, and they are both believed to exhibit spontaneous ``chiral'' symmetry\footnote{There is no chirality for fermions in three dimensions.  The symmetry in question would be a chiral symmetry if the same theory were considered in four dimensions.} breaking when the number of fermion flavors in QED$_3$ is sufficiently small \cite{Pisarski:1984dj, Appelquist:1988sr, Nash:1989xx}.  At the same time, QED$_3$ is in many ways simpler than QCD$_4$.  While QCD$_4$ is a non-abelian gauge theory that requires non-trivial renormalization and in which asymptotic freedom follows from a complicated beta-function computation \cite{Gross:1973id, Politzer:1973fx}, QED$_3$ is a super-renormalizable abelian gauge theory where asymptotic freedom follows from dimensional analysis.  In three dimensions, the gauge coupling $e^2$ has units of mass, and hence the dimensionless coupling $e^2 / \mu$, $\mu$ being the renormalization group (RG) scale, becomes arbitrarily small in the ultraviolet.  Therefore, understanding the dynamics of QED$_3$ should be in principle less onerous than understanding that of QCD$_4$, and at the same time QED$_3$ may provide us with valuable lessons about the physics that leads to spontaneous chiral symmetry breaking.

To be precise, let $N_f$ be the number of two-component complex fermions in QED$_3$.  The following discussion will be restricted to the case where $N_f$ is even, because otherwise it would be impossible to regularize the theory while preserving parity and time reversal symmetry.  In Euclidean signature, the action is
 \es{ActionEuc}{
  {\cal S} = \int d^3r \, \left[\frac{1}{4 e^2} F_{\mu\nu} F^{\mu\nu} +  \sum_{a=1}^{N_f} \psi_a^\dagger \left(i \slashed{D} + \slashed{A} \right) \psi_a \right] \,,
 }
where $\psi_a$ are the fermion fields, $A_\mu$ is a $U(1)$ gauge field with field strength $F_{\mu\nu}$, and $\slashed{D}$ is the Dirac operator.   This theory has a global $SU(N_f)$ symmetry under which the fermions $\psi_a$ transform as a fundamental vector.  Chiral symmetry breaking refers to the breaking of this symmetry to $SU(N_f/2) \times SU(N_f/2) \times U(1)$.  While the maximal value $N_f^\text{crit}$ for which one expects this symmetry breaking to occur has been the subject of some debate, recent lattice gauge theory results \cite{Hands:2004bh, Strouthos:2008kc, Strouthos:2008hs} suggest that $N_f^\text{crit} = 2$.\footnote{Other approaches give higher critical values $N_f^\text{crit}$.  For instance, an analytical approach based on solving the Schwinger-Dyson equations self-consistently gives $N_f^\text{crit} = 6$ \cite{Maris:1996zg, Fischer:2004nq, Goecke:2008zh}.  Bounds obtained by using the $F$-theorem \cite{Myers:2010tj, Jafferis:2011zi, Klebanov:2011gs, Casini:2012ei} give $N_f^\text{crit} \leq 6$ \cite{Grover:2012sp}.}  For $N_f > 2$ the theory is believed to flow to a strongly interacting conformal field theory (CFT) in the infrared.  At least at large enough $N_f$, the CFT is obtained by simply taking the limit $e^2 \to \infty$ in the action \eqref{ActionEuc}.

To a condensed matter physicist, QED$_3$ may be of interest because it describes, for instance, the effective low-energy dynamics of ``algebraic spin liquids'' \cite{Rantner:2002zz}.  As explained in \cite{Wen:2001yp}, $SU(N_f)$ spins on a lattice can be described in a slave fermion formalism \cite{Affleck:1987zz} as compact QED$_3$ with $N_f$ fermion flavors.  Compactness of the $U(1)$ gauge field means that monopole configurations are not suppressed.   Just like in the case of no fermions studied by Polyakov \cite{Polyakov:1975rs, Polyakov:1976fu}, at small $N_f$ one can argue that these monopoles proliferate, and their proliferation leads to confinement of electric charges.  In renormalization group language, the monopoles can proliferate provided that the operators that create them are relevant in the RG sense \cite{Hermele, alicea}.  When $N_f$ is large, it can be shown that the monopole operators become irrelevant \cite{Borokhov:2002ib}, and therefore one can ignore the compactness of the gauge group \cite{Hermele}.  All other relevant operators, such as fermion bilinears and Chern-Simons interactions, are suppressed provided that one requires invariance under various discrete symmetries, in which case the resulting low-energy effective theory, namely the CFT obtained by taking $e^2 \to \infty$ in \eqref{ActionEuc}, is a stable RG fixed point. This stable CFT is the low-energy effective theory of an algebraic spin liquid.

In this paper, I will calculate the scaling dimension of the monopole operator that inserts one unit of magnetic flux by performing an expansion to next-to-leading order in $1/N_f$.  The leading term in this expansion was found by Borokhov, Kapustin, and Wu~\cite{Borokhov:2002ib}.  I will therefore improve on their result.  While the validity of the $1/N_f$ expansion has not been tested in QED$_3$, there is evidence that in supersymmetric versions of QED$_3$ the $1/N_f$ expansion gives reasonably accurate results even for small values of $N_f$ \cite{Klebanov:2011td, Safdi:2012re}.  The tests of the $1/N_f$ expansion were made possible in supersymmetric theories by the technique of supersymmetric localization \cite{Pestun:2007rz, Kapustin:2009kz} combined with $F$-maximization \cite{Jafferis:2010un, Closset:2012vg}, whereby one can obtain exact results for various supersymmetry-protected quantities, even at strong coupling.  It is therefore desirable to make use of the $1/N_f$ expansion in non-supersymmetric theories as well, and this paper will focus on QED$_3$.

In QED$_3$, the scaling dimensions of the operators in the zero-monopole sector can be computed in the $1/N_f$ expansion using Feynman diagrams, and many results are known to several orders in $1/N_f$ (see, for example, \cite{Rantner:2002zz}).  The situation is significantly harder when one includes monopoles, because the monopole creation operators do not have simple expressions in terms of the fundamental fields of the theory \eqref{ActionEuc}, so conventional Feynman diagrams are not of much use.  The monopole scaling dimensions can be computed, however, using the state-operator correspondence,\footnote{For other instances where the scaling dimensions of monopole operators were computed using the state-operator correspondence, see \cite{Borokhov:2002ib, Borokhov:2002cg, Metlitski:2008dw, Pufu:2013eda}.  See also \cite{Murthy:1989ps} for a different method of computing scaling dimensions of monopole operators.} which maps local operators of a CFT inserted at the origin of $\R^3$ to normalizable states on $S^2 \times \R$, the $\R$ coordinate being interpreted as Euclidean time.  The scaling dimension on $\R^3$ is mapped to the energy of the state on $S^2$.

A monopole operator of strength $\myn$ is a local operator that changes the boundary condition for the gauge field at the insertion point such that the field strength $F = dA$ integrates to 
 \es{FInt}{
  \int_{S^2} F = 2 \pi \myn 
 }
over any (sufficiently small) two-sphere surrounding the insertion point.  Here, Dirac quantization imposes $\myn \in \Z$.  The corresponding state on $S^2$ is the ground state in the presence of $\myn$ units of magnetic flux through $S^2$;  the excited states in the presence of this magnetic flux correspond to composite operators that contain a monopole operator of strength $\myn$.  Standard thermodynamics equates the ground state energy with the free energy on $S^2$, which can be computed as minus the logarithm of the $S^2 \times \R$ partition function.  The scaling dimension of a monopole operator of strength $\myn$ can therefore be extracted from the $S^2 \times \R$ partition function in the background of $\myn$ units of magnetic flux through the $S^2$.

The partition function on  $S^2 \times \R$ can be evaluated perturbatively in $1/N_f$ as follows. To leading order in $N_f$ one can ignore the fluctuations of the gauge field and perform a Gaussian integral over the fermionic fields.  The results obtained by this procedure agree with those of \cite{Borokhov:2002ib}.  The next order term in the large $N_f$ expansion comes from performing a functional integral over the fluctuations of the gauge field.  This integral is also Gaussian, because the higher order terms in the effective action for the gauge field fluctuations are suppressed by positive powers of $1/N_f$. The evaluation of this Gaussian integral is the main focus of this paper.   Similar but significantly simpler computations were performed in \cite{Pufu:2013eda}, where the contribution to the $S^2$ ground state energy coming from a scalar fluctuation was computed in a similar setup, and in \cite{Klebanov:2011td}, where an integral over gauge field fluctuations on $S^3$ was computed in the absence of any monopoles.

In contrast with the computation presented in this paper, a similar computation in a supersymmetric theory would be much simpler if, as mentioned above, one appeals to supersymmetric localization and $F$-maximization \cite{Jafferis:2010un}.  Indeed, in ${\cal N} =2$ supersymmetric QED with $N$ chiral super-fields with charge $+1$ and $N$ chiral super-fields with charge $-1$ under the $U(1)$ gauge group, the scaling dimension of a BPS monopole operator with strength $\myn=1$ is $N(1 - \Delta)$ \cite{Benini:2009qs, Benini:2011cma}, where $\Delta$ is the scaling dimension (or R-charge) of one of the chiral super-fields.  This scaling dimension was computed using $F$-maximization in \cite{Klebanov:2011td} both exactly and in a $1/N$ expansion, and the two calculations agree very well even at small values of $N$---see Figure~2 in \cite{Klebanov:2011td}.  The scaling dimension of the $\myn=1$ monopole operator  is in this case
 \es{MonSuper}{
  \frac N2 + \frac{2}{\pi^2} + \frac{2 \left(\pi^2 - 12 \right)}{\pi^4 N} + O(N^{-2}) \,.
 }
In the non-supersymmetric case presented below, there are no methods that yield a simple analytic answer like \eqref{MonSuper}, and one has to resort instead to numerical methods.

The remainder of this paper is organized as follows. Section~\ref{CONVENTIONS} contains the setup of the problem in more detail.  In Section~\ref{LEADING}, I reproduce the leading order results of \cite{Borokhov:2002ib} by performing the Gaussian integral over the fermionic fields.  In Section~\ref{STRATEGY}, I outline the strategy used for computing the leading $1/N_f$ correction to these results.  Section~\ref{GREEN} contains expressions for the fermion Green's function on $S^2 \times \R$ in the presence of $\myn$ units of magnetic flux through $S^2$.  This Green's function is the central ingredient  in the effective action for the gauge field fluctuations.  In Section~\ref{KERNEL}, I evaluate the Gaussian integral over the gauge field fluctuations.  Lastly, Section~\ref{DISCUSSION} contains a discussion of these results.

\section{Setup and conventions}
\label{CONVENTIONS}

On an arbitrary conformally flat Riemannian three-manifold, the QED$_3$ action with $N_f$ two-component complex fermions is
 \es{Action}{
  {\cal S} = \sum_{a=1}^{N_f} \int d^3 r\, \sqrt{g} \left[ \psi_a^\dagger \left(i \slashed{D} + \slashed{\cal A} + \slashed{A} \right) \psi_a  \right] \,,
 }
where $g$ is the determinant of the metric, $\psi_a$ are the $N_f$ fermion fields, and the gauge field is written as the sum of a background ${\cal A}_\mu$ and a small fluctuation $A_\mu$ around this background.  We are interested in studying this theory on $S^2 \times \R$ in the background of $\myn$ units of magnetic flux through $S^2$.  Parameterizing $S^2 \times \R$ using coordinates $r = (\theta, \phi, \tau)$ such that the metric is written as
 \es{MetricS2R}{
  ds^2  = d\theta^2 + \sin^2 \theta d \phi^2 + d\tau^2 \,,
 }
one can take the background gauge field to be
 \es{BackgroundGauge}{
  {\cal A}(r) = \frac {\myn} 2 \left(1 - \cos \theta \right) d \phi\,.
 }
The field strength ${\cal F} = d {\cal A}$ integrates to $2 \pi n$ over $S^2$, as in \eqref{FInt}.  The expression \eqref{BackgroundGauge} is well-defined everywhere away from the South pole at $\theta = \pi$, where there is a Dirac string extended in the $\R$ direction.  The requirement that this Dirac string should be invisible restricts $\myn \in \Z$.  We will assume $\myn \geq 0$.

In working with spinors on a curved manifold, one should specify the conventions used for the frame and gamma matrices.  To simplify the subsequent analysis, we can introduce the frame obtained from conformal transformation of the standard one in $\R^3$:
 \es{Frame}{
  e^i = e^{-\tau} dx^i \,,
   \qquad
    \vec{x} = e^\tau \, \hat x =  e^\tau \begin{pmatrix}
     \sin \theta \cos \phi & \sin \theta \sin \phi & \cos \theta 
    \end{pmatrix} \,,
 }
which can be found by writing the standard line element on $\R^3$ in spherical coordinates as $d\vec{x}^2 = d\rho^2 + \rho^2 (d\theta^2 + \sin^2 \theta d \phi^2)$ such that the metric \eqref{MetricS2R} on $S^2 \times \R$ is
 \es{Metric}{
  ds^2 = \frac{d\vec{x}^2}{\rho^2} \,, \qquad \rho = e^\tau \,.
 }
We will use the gamma matrices $\gamma_i = \sigma_i$ where $\sigma_i$ are the Pauli matrices.\footnote{There is no difference between upper and lower frame indices in Euclidean signature.}

The following sections are devoted to the computation of the ground state energy $F^{(\myn)} = - \log Z^{(\myn)}$ on $S^2 \times \R$ in the presence of the background \eqref{BackgroundGauge}.   This quantity can be expanded at large $N_f$ as
 \es{Expansion}{
  F^{(\myn)} = N_f F^{(\myn)}_0 + F^{(\myn)}_1 + \ldots \,.
 }
One expects $F^{(0)} = 0$ because when $n=0$ the ground state on $S^2$ corresponds to the identity operator on $\R^3$, which has vanishing scaling dimension.  While we will check explicitly that $F^{(\myn)}_0 = 0$, we will take $F^{(0)}_1  = 0$ as an assumption and identify $F^{(\myn)}_1$ with $F^{(\myn)}_1 - F^{(0)}_1$.

\section{Leading order free energy}
\label{LEADING}

To leading order in $N_f$ we can ignore the fluctuations of the gauge field, and the action becomes that of free fermions in the background gauge field \eqref{BackgroundGauge}.   This action can be written in almost diagonal form by expanding the fermion fields in terms of Fourier modes in the $\R$ direction as well as in terms of analogs of the $S^2$ spherical harmonics that are appropriate for describing a spin-$1/2$ charged particle in the monopole background \eqref{BackgroundGauge}.  The main ingredients in constructing these harmonics are the monopole spherical harmonics $Y_{\myn/2, \ell m}$ \cite{Wu:1977qk,Wu:1976ge}, with $\ell \geq \ell/2$ and $-\ell \leq m \leq \ell$, which are simultaneous eigenfunctions of the gauge-covariant angular momentum operators:
 \es{ScalarHarm}{
  \vec{L}^2 Y_{\myn/2, \ell m} = \ell(\ell+1) Y_{\myn/2, \ell m} \,, 
  \qquad
   L_3 Y_{\myn/2, \ell m} = m Y_{\myn/2, \ell m} \,.
 }
For explicit formulas for $Y_{\myn/2, \ell m}$ in terms of the angles on $S^2$, see \cite{Wu:1977qk,Wu:1976ge} or Appendix~\ref{HARMONICS}.  The monopole spherical harmonics form a complete basis of functions on $S^2$ suited for describing {\em spinless} particles with unit gauge charge in the background of the monopole \eqref{BackgroundGauge}.  When $\myn=0$ they are nothing but the usual $S^2$ spherical harmonics.  

As explained in more detail in Appendix~\ref{HARMONICS}, the scalar harmonics $Y_{\myn/2, \ell m}$ can be generalized to include spin by simultaneously diagonalizing $\vec{S}^2$, $\vec{L}^2$, ${\vec J}^2$, and $J_3$, where $\vec{S}$, $\vec{L}$, and $\vec{J}$ are the spin, orbital, and total angular momentum operators.  The spin-$1/2$ case yields two sets of spinors that we denote by $S_{\myn, \ell m}$ and $T_{\myn, \ell m}$.  They have orbital angular momentum quantum number $\ell$ and total angular momentum quantum numbers
 \es{QuantumNumbers}{
  T_{\myn, \ell m}: \qquad
    j = \ell + \frac 12 \,, \qquad m_j = m + \frac 12 \,, \\
  S_{\myn, \ell m}: \qquad
    j = \ell - \frac 12 \,, \qquad m_j = m + \frac 12 \,.  
 } 

The expansion of the fermion fields reads 
  \es{psiExpansion}{
  \psi_a(r) = \int \frac{d\omega}{2 \pi} \left[  \sum_{\ell = \frac{\myn}{2}}^\infty \sum_{m = -\ell-1}^\ell \Psi_{a, T}^{\ell m} (\omega) T_{\myn, \ell m}(\theta, \phi) 
    + \sum_{\ell = \frac{\myn}{2}}^\infty \sum_{m = -\ell}^{\ell-1}  \Psi_{a, S}^{\ell m} (\omega) S_{\myn, \ell m}(\theta, \phi) 
    \right] e^{-i \omega \tau}  \,,
 }
where $\Psi_{a, T}^{\ell m}$ and $\Psi_{a, S}^{\ell m}$ are (anti-commuting) coefficients.  In this expression and throughout this section all spinor indices carried by $\psi_a$, $S_{\myn, \ell m}$, and $T_{\myn, \ell m}$ are suppressed.   The range of $m$ in the sums in \eqref{psiExpansion} follows from \eqref{QuantumNumbers} and the usual $-j \leq m_j \leq j$.  While the spin-$1/2$ monopole harmonics $S_{\myn, \ell m}$ and $T_{\myn, \ell m}$ diagonalize $\vec{S}^2$, $\vec{L}^2$, $\vec{J}^2$, and $J_3$, they are not eigenspinors of the gauge-covariant Dirac operator $i \slashed{D} + \slashed{\cal A}$ because this operator does not commute with $\vec{L}^2$. Indeed, for given $j = \ell - 1/2$ and $m_j = m + 1/2$, there are only two spinors that differ in their $\vec{L}^2$ quantum number, namely $T_{\myn, (\ell-1)m}$ and $S_{\myn, \ell m}$;  starting from \cite{Borokhov:2002ib} it can be shown that
 \es{DerBasis}{
  \left(i \slashed{D} + \slashed{\cal A} \right)
  \begin{pmatrix} T_{\myn, (\ell-1)m} e^{-i \omega \tau} \\ S_{\myn, \ell m} e^{-i \omega \tau} \end{pmatrix}
   = {\bf N}_{\myn, \ell} \left( \omega + i {\bf M}_{\myn, \ell} \right)
    \begin{pmatrix} T_{\myn, (\ell-1)m}  e^{-i \omega \tau} \\ S_{\myn, \ell m} e^{-i \omega \tau} \end{pmatrix} \,,
 }
where the matrices ${\bf M}_{\myn, \ell}$ and ${\bf N}_{\myn, \ell}$ are given by\footnote{I thank Mark Mezei for helping me correct a sign error in a previous version of this equation.} 
 \es{GotM}{
  {\bf M}_{\myn, \ell} = \begin{pmatrix}
   \ell \left(1- \frac{\myn^2}{4\ell^2} \right) & -\frac{\myn}{2} \sqrt{1 - \frac{\myn^2}{4\ell^2}} \\
   -\frac{\myn}{2} \sqrt{1 - \frac{\myn^2}{4\ell^2}} & -\ell \left(1 - \frac{\myn^2}{4\ell^2} \right) 
  \end{pmatrix} \,, \qquad
   {\bf N}_{\myn, \ell} = \begin{pmatrix}
    -\frac{n}{2\ell} & -\sqrt{1 - \frac{n^2}{4\ell^2}} \\
    -\sqrt{1 - \frac{\myn^2}{4\ell^2}} & \frac{n}{2\ell}
   \end{pmatrix} \,.
 }  
Note that these matrices act trivially on the spinor indices of  $T_{\myn, (\ell-1)m}$ and $S_{\myn, \ell m}$, which, as mentioned above, are consistently being suppressed.

The case $\ell = \myn/2$ deserves a comment.  Since there are no scalar monopole spherical harmonics with orbital angular momentum less than $\myn/2$, when $\ell = \myn/2$ (or equivalently when $j = (\myn-1)/2$) the matrices in \eqref{GotM} should be thought of as $1 \times 1$ matrices equal to the bottom right entries of the expressions in \eqref{GotM}.  In particular, ${\bf M}_{\myn, \myn/2} = 0$ and ${\bf N}_{\myn, \myn/2} = 1$.  The modes $S_{\myn, (\myn/2)m}$ are the $\myn$ zero modes of the Dirac operator on $S^2$ in the presence of $\myn$ units of magnetic flux.

Using \eqref{DerBasis} and \eqref{psiExpansion} and ignoring the gauge field fluctuations, the action takes the block-diagonal form
 \es{EffAction}{
  {\cal S}_0 = \sum_{a=1}^{N_f} \int \frac{d\omega}{2 \pi} \sum_{\ell = \frac{\myn}{2}}^\infty \sum_{m = -\ell}^{\ell-1} 
   \begin{pmatrix}
    \Psi_{a, T}^{(\ell-1) m}(\omega)^* & \Psi_{a, S}^{\ell m}(\omega)^*
   \end{pmatrix}
    {\bf N}_{\myn, \ell} \left(\omega + i {\bf M}_{\myn, \ell} \right)
    \begin{pmatrix}
     \Psi_{a, T}^{(\ell-1) m}(\omega) \\
     \Psi_{a, S}^{\ell m}(\omega) 
    \end{pmatrix} \,.
 }
After performing the Gaussian integral over the fermions, the logarithm of the $S^2 \times \R$ partition function is thus
 \es{LeadingFree}{
   \log Z^{(n)} = N_f \int \frac{d \omega}{2 \pi} \sum_{\ell = \frac{\myn}{2}}^\infty 2\ell \log \det \left( {\bf N}_{\myn, \ell} \left(\omega + i {\bf M}_{\myn, \ell} \right) \right) \,,
 }
the factor of $2 \ell$ coming from the sum over $m$, or equivalently from the $2j+1 = 2\ell$ degeneracy.  The determinant in \eqref{LeadingFree} is easily computed, and the coefficient $F_0^{(n)}$ in \eqref{Expansion} can be expressed as
 \es{LeadingF}{
  F_0^{(n)} = -\int \frac{d \omega}{2 \pi} \sum_{\ell = \frac{\myn}{2}}^\infty 2\ell \log \left(\omega^2  + \ell^2 - \frac {\myn^2}{4} \right)  \,.
 }
This expression is divergent and requires regularization.  One way of extracting the finite part is to write $\log A = - dA^{-s} / ds \Bigr|_{s=0}$, followed by an evaluation of the sum and integral in \eqref{LeadingF} at values of $s$ where they are absolutely convergent, and then by an analytic continuation of the answer to $s = 0$.  (See also \cite{Pufu:2013eda}, where a similar expression was regularized in the same fashion.)  Performing the $\omega$ integral yields
 \es{LeadingFAgain}{
  F_0^{(n)} = -\sum_{\ell = \frac{\myn}{2}}^\infty 2 \ell \left( \ell^2 - \frac {\myn^2}{4} \right)^{\frac 12 - s} \Biggr|_{s=0}  \,.
 }
After adding and subtracting quantities that are divergent when $s=0$, one can write this sum as
 \es{F0Reg}{
  F_0^{(n)} &= -2\sum_{\ell = \frac{\myn}{2}}^\infty \left[ \ell \left( \ell^2 - \frac {\myn^2}{4} \right)^{\frac 12 - s}
     - \ell^{2-2s} + \frac{\myn^2 }{8}  (1-2s) \ell^{-2s} \right]  \Biggr|_{s=0} \\
   &\qquad\qquad\qquad\qquad\qquad\qquad {}+ 2 \sum_{\ell = \frac{\myn}{2}}^\infty \left[-\ell^{2-2s} + \frac{\myn^2}{8}(1-2s) \ell^{-2s} \right]  \Biggr|_{s=0} \,.
 }
 The first sum is absolutely convergent, so one can simply set $s=0$ and evaluate it numerically.  The second sum can be evaluated using zeta-function regularization.  The result is
  \es{F0RegFinal}{
    F_0^{(n)} = -2\sum_{\ell = \frac{\myn}{2}}^\infty \left[ \ell \sqrt{\ell^2 - \frac {\myn^2}{4} } - \ell^2 + \frac{\myn^2}{8} \right]
   -\frac{n(n-1)(n+4)}{24} \,.
  }  
 See Table~\ref{F0Table} for $F_0^{(n)}$ evaluated for the first few lowest values of $\myn$.  
  \begin{table}[t!]
%\caption{default}
\begin{center}
\begin{tabular}{c|c}
  $\myn$ & $F_0^{(n)} $ \\
  \hline \hline
  $0$ & $0$ \\
  $1$ & $0.265$ \\
  $2$ & $0.673$ \\
  $3$ & $1.186$ \\
  $4$ & $1.786$ \\
  $5$ & $2.462$ \\
  $6$ & $3.206$ 
\end{tabular}
\end{center}
\caption{The values of $F_0^{(n)}$ as obtained in a few particular cases by evaluating \eqref{F0RegFinal}.}
\label{F0Table}
\end{table}%
 These results agree with those of \cite{Borokhov:2002ib}, where the sum \eqref{LeadingFAgain} was regularized by a different method.

\section{Beyond leading order:  The strategy}
\label{STRATEGY}

The $1/N_f$ corrections to the ground state energy on $S^2 \times \R$ come from performing the functional integral over the gauge field.  The effective action for the gauge field fluctuations obtained after integrating out the fermions is
 \es{EffActionGauge}{
  {\cal S}_{A} = \frac 12 \int d^3 r\, d^3 r'\, \sqrt{g(r)} \sqrt{g(r')} A_i(r) K^{n, ij}(r, r') A_j(r') + \ldots \,,
 }
where the ellipsis denotes terms that are higher order in the gauge field fluctuations, and $K^{n, ij}(r, r')$ is the two-point correlator of the $U(1)$ current $J^i$:
 \es{KDef}{
  K^{n, ij} (r, r') = -\langle J^i(r) J^j(r') \rangle_n \,.
 }
Here, $i$ and $j$ denote frame indices in the frame defined in \eqref{Frame}.  The subscript $\myn$ on the angle brackets serves as a reminder that the expectation value should be evaluated in the monopole background \eqref{BackgroundGauge}.  As can be read off from the action \eqref{Action}, the current $J^i$ is
 \es{Current}{
  J^i = \sum_{a=1}^{N_f}  \psi_a^\dagger \sigma^i \psi_a \,.
 }
In computing the two point function \eqref{KDef} we should treat the $N_f$ fermions as non-interacting, so
 \es{KDefAgain}{
  K^n_{ij} (r, r') = -N_f \langle \psi^\dagger(r) \sigma_i \psi(r) \psi^\dagger(r') \sigma_j \psi(r') \rangle   \,,
 }
where $\psi$ corresponds to a single charged fermion in the background gauge field given by \eqref{BackgroundGauge}.   In terms of the fermion Green's function $G_n(r, r') = \langle \psi(r) \psi^\dagger(r') \rangle_n$, one can write the kernel \eqref{KDefAgain} as 
 \es{KqKernelAgain}{
  K^n_{ij}(r, r') = -N_f \tr \left( \sigma_i G_n(r, r') \sigma_j G_n^\dagger(r, r') \right) \,.
 }

The reason why one can ignore the higher order terms in \eqref{EffActionGauge} now emerges.  The factor of $N_f$ can be absorbed by rescaling the gauge field fluctuations $A_i \to A_i / \sqrt{N_f}$.  After this rescaling, the higher order terms in the gauge field fluctuations that were omitted from \eqref{EffActionGauge} become suppressed at large $N_f$, so the leading contribution to the $S^2 \times \R$ free energy comes from the Gaussian integral over the gauge field fluctuations computed using just the quadratic action \eqref{EffActionGauge}.  

The action \eqref{EffActionGauge} can be written in diagonal form by expanding both the fluctuations $A_i$ and the kernel $K^n_{ij}$ in plane waves in the $\R$ coordinate and vector spherical harmonics on $S^2$.  Just as in the case of the spinor harmonics discussed in the previous section, the vector harmonics can be defined by simultaneously diagonalizing the angular momentum operators $\vec{S}^2$, $\vec{L}^2$, $\vec{J}^2$, and $J_3$.  See Appendix~\ref{HARMONICS} for more details.  These vector harmonics are constructed from the usual $S^2$ spherical harmonics $Y_{\ell m}$ because the gauge field does not experience monopole flux.  They have spin $s = 1$, orbital angular momentum $\ell$, and total angular momentum quantum numbers given by
 \es{VectorQuantum}{
  &U_{\ell m}^i: \qquad j = \ell - 1 \geq 0 \,, \qquad m_j = m \,, \\
  &V_{\ell m}^i: \qquad j = \ell \geq 1 \,, \qquad m_j = m \,, \\
  &W_{\ell m}^i: \qquad j = \ell + 1 \geq 1\,, \qquad m_j = m \,.
 }

We thus expand the gauge field fluctuations as
 \es{GaugeExpansion}{
  A^i(r) &= \int \frac{d\omega}{2 \pi} 
   \Biggl[ \sum_{\ell=1}^\infty \sum_{m=-\ell+1}^{\ell-1} a_U^{\ell m}(\omega)  U^i_{\ell m}(\theta, \phi) 
    + \sum_{\ell=1}^\infty \sum_{m=-\ell}^\ell a_V^{\ell m}(\omega)  V^i_{\ell m}(\theta, \phi) \\
    &\qquad\qquad\qquad\qquad\qquad\qquad\qquad\qquad{}+ 
    \sum_{\ell=0}^\infty \sum_{m=-\ell-1}^{\ell+1} a_W^{\ell m}(\omega)  W^i_{\ell m}(\theta, \phi)  \Biggr] e^{-i \omega \tau} \,,
 }
where the summation ranges follow from \eqref{VectorQuantum}.   The kernel $K^\myn_{ij}(r, r')$, seen as a $3\times 3$ matrix ${\bf K}^{\myn}(r, r')$, should also be expanded in terms of the vector harmonics as
 \es{KExpansion}{
  {\bf K}^{n} (r, r') = \int \frac{d\omega}{2 \pi} \sum_{\ell = 0}^\infty \sum_{m = -\ell}^\ell 
   e^{-i \omega (\tau - \tau')} 
   \begin{pmatrix}
    U_{(\ell+1)m}(\theta, \phi) & V_{\ell m}(\theta, \phi) & W_{(\ell-1)m}(\theta, \phi)
   \end{pmatrix} \\
    \times {\bf K}^n_\ell(\omega) 
   \begin{pmatrix}
    U_{(\ell+1)m}^{\dagger}(\theta', \phi') \\ V_{\ell m}^{\dagger}(\theta', \phi') \\ W_{(\ell-1)m}^{\dagger}(\theta', \phi')
   \end{pmatrix} \,,
 }
where, if $\ell>0$, ${\bf K}^n_\ell(\omega)$ is a $3 \times 3$ hermitian matrix whose entries depend on $\ell$ and $\omega$.  When $\ell=0$, $V_{\ell m}$ and $W_{(\ell-1)m}$ do not exist, and this matrix should be thought of as $1 \times 1$.   Note that the matrix ${\bf K}^n_\ell(\omega)$ acts trivially on the frame indices of the harmonics and can be computed by inverting \eqref{KExpansion}:
 \es{GotKq}{
  {\bf K}^n_\ell(\omega)  = \frac{4 \pi}{2\ell+1} \int d^3 r\, \sqrt{g(r)}
   e^{i \omega \tau} 
    \sum_{m=-\ell}^\ell \begin{pmatrix}
    U^\dagger_{(\ell+1)m}(\theta, \phi) & V^\dagger_{\ell m}(\theta, \phi) & W^\dagger_{(\ell-1)m}(\theta, \phi)
   \end{pmatrix} \\
    \times {\bf K}^n(r, r')
   \begin{pmatrix}
    U_{(\ell+1)m}(\theta', \phi') \\ V_{\ell m}(\theta', \phi') \\ W_{(\ell-1)m}(\theta', \phi')
   \end{pmatrix} \Biggr\rvert_{r' = 0} \,,
 }
where we made use of the symmetry under $S^2$ rotations and translations along $\R$.  Using \eqref{GaugeExpansion} and \eqref{KExpansion}, one can now write down the effective action \eqref{EffActionGauge} in almost diagonal form:
 \es{GaugeDiagonal}{
  {\cal S}_A = \int \frac{d\omega}{2 \pi} \sum_{\ell=1}^\infty \sum_{m=-\ell}^\ell 
   \begin{pmatrix}
    a_U^{\ell m}(\omega)^* & a_V^{\ell m}(\omega)^* & a_W^{\ell m}(\omega)^*
   \end{pmatrix}
    {\bf K}^n_\ell(\omega)
   \begin{pmatrix}
    a_U^{\ell m}(\omega) \\ a_V^{\ell m}(\omega) \\ a_W^{\ell m}(\omega)
   \end{pmatrix} \,.
 }
The Gaussian integral over the gauge field fluctuations then gives, roughly,
 \es{FRough}{
  F^{(n)}_0 = \frac 12 \int \frac{d\omega}{ 2\pi} \sum_{\ell=0}^\infty (2 \ell+1) \log \det {\bf K}_\ell^n(\omega) \,.
 }
This expression is rough because it ignores a very important subtlety:  gauge invariance.  Indeed, the effective action ${\cal S}_A$ should be independent of the pure gauge modes that are part of the expansion \eqref{GaugeExpansion}, so it must be true that the matrix ${\bf K}_\ell^n(\omega)$ has many eigenvalues equal to zero.  (To be precise, for each $\ell$ there should be one eigenvalue that vanishes.) While it should be possible to work carefully in a fixed gauge, the subtleties related to gauge fixing are exactly the same for all $\myn$ and they disappear from the differences\footnote{The sum starts at $\ell=1$ because, as mentioned above, when $\ell=0$ the matrix ${\bf K}_\ell^n(\omega)$ has only one entry, which vanishes for all $n$ because it corresponds to a pure gauge mode.}
 \es{FDiff}{
  F^{(n)}_0 - F^{(0)}_0 = \frac 12 \int \frac{d\omega}{ 2\pi} \sum_{\ell=1}^\infty (2 \ell+1) \log \det 
    \frac{ {\bf K}_\ell^n(\omega) }{ {\bf K}_\ell^0(\omega) } \,,
 }
provided that we only take the ratio of the non-zero eigenvalues of these matrices. One can furthermore assume that $F^{(0)}_0 = 0$, because when $n=0$ the $S^2$ ground state energy equals the scaling dimension of the identity operator, which vanishes.

It can be checked explicitly using the formulas in the following sections that $K^{n, UV}_\ell(\omega) = K^{n, VW}_\ell(\omega) = 0$, which implies that the matrix ${\bf K}^{n}_\ell(\omega)$ takes the form
 \es{KSimple}{
  {\bf K}^n_\ell(\omega) = \begin{pmatrix}
   K^{n, UU}_\ell(\omega) & 0 & K^{n, UW}_\ell(\omega) \\
   0& K^{n, VV}_\ell(\omega) & 0 \\
   K^{n, WU}_\ell(\omega) & 0 & K^{n, WW}_\ell(\omega) 
  \end{pmatrix} \,.
 }
The entry $K^{n, VV}_\ell(\omega)$ is a (non-vanishing) eigenvalue of this matrix.  Since one other eigenvalue vanishes and since the trace of this matrix is the sum of all three eigenvalues, the third eigenvalue must be $K^{\myn, UU}_\ell + K^{\myn, WW}_\ell$.  Let's denote the non-zero eigenvalues by
 \es{NonZeroEvalues}{
  K^{\myn, E}_\ell \equiv K^{\myn, UU}_\ell + K^{\myn, WW}_\ell\,, \qquad  K^{\myn, B}_\ell \equiv K^{\myn, VV}_\ell \,.
 }
The labels $E$ and $B$ stand for ``$E$-modes'' and ``$B$-modes,'' respectively, because when restricted to $S^2$ one can check using \eqref{UVWExplicit} that $U_{\ell m}$ and $W_{\ell m}$ have vanishing curl (just like an electric field), while $V_{\ell m}$ has vanishing divergence (just like a magnetic field).

To sum up, the correction to the free energy obtained after performing the Gaussian integral over the gauge field fluctuations is
 \es{FCorrected}{
   F^{(n)}_1 = \frac 12 \int \frac{d\omega}{2 \pi} \sum_{\ell=1}^\infty (2 \ell+1) \log \frac{K_\ell^{n, E}(\omega) K_\ell^{n, B}(\omega)}{K_\ell^{0, E}(\omega) K_\ell^{0, B}(\omega)} \,.
 }
Explicit formulas for $K_\ell^{n, E}(\omega)$ and $K_\ell^{n, B}(\omega)$ will be given in Section~\ref{KERNEL}.  To find them, one should start by computing the fermion Green's function needed in \eqref{KqKernelAgain}, which is the goal of the following section.

\section{Fermion Green's function}
\label{GREEN}

The Green's function in the background \eqref{BackgroundGauge} is defined as 
 \es{GreenDef}{
  G_\myn (r, r') = \langle \psi (r) \psi^\dagger(r') \rangle_\myn \,.
 }
It satisfies $\left(i \slashed{D} + \slashed{\cal A} \right)(r) G_\myn(r,r') = -\delta(r - r')$, so we can write it using the spectral decomposition as
 \es{Spectral}{
  G_\myn(r, r') = -\int \frac{d\omega}{2 \pi} \sum_{\ell = \frac{\myn}2}^\infty \sum_{m = -\ell}^{\ell-1} 
   \begin{pmatrix}
    T_{\myn, (\ell-1) m}(\theta, \phi) & S_{\myn, \ell m}(\theta, \phi)
   \end{pmatrix}
    \frac{e^{-i \omega(\tau - \tau')}}{{\bf N}_{\myn, \ell} \left(\omega + i {\bf M}_{\myn, \ell}\right) }
    \begin{pmatrix}
     T_{\myn, (\ell-1) m}^\dagger(\theta', \phi') \\
     S_{\myn, \ell m}^\dagger(\theta', \phi')
    \end{pmatrix} \,,
 }
where the matrices ${\bf M}_{\myn, \ell}$ and ${\bf N}_{\myn, \ell}$ were introduced in \eqref{GotM}.  Using \eqref{DerBasis}, it is not hard to see that acting with $\left(i \slashed{D} + \slashed{\cal A} \right)(r)$ on the Green's function results in a delta function.

To use this expression further, note that
 \es{InverseMatrix}{
  \frac{1}{\omega + i {\bf M}_{\myn, \ell}} = \frac{\omega - i {\bf M}_{\myn, \ell} }{\omega^2 + \ell^2 - n^2/4} \,, \qquad \frac{1}{{\bf N}_{\myn, \ell}} = {\bf N}_{\myn, \ell} \,.
 }
One can perform the $\omega$ integral:
 \es{SpectralAgain}{
  G_\myn(r, r') &=  \sum_{\ell = \frac{\myn}{2}}^\infty \sum_{m = -\ell}^{\ell-1} \frac{i e^{-E_{\myn, \ell} \abs{\tau - \tau'}}}{2}
   \begin{pmatrix}
    T_{\myn, (\ell-1) m}(\theta, \phi) & S_{\myn, \ell m}(\theta, \phi)
   \end{pmatrix} \\
   &\qquad\qquad\qquad\times \left[\sgn (\tau - \tau') {\bf N}_{\myn, \ell}  + \begin{pmatrix}
    0 & -1 \\
    1 & 0
   \end{pmatrix} \right] 
    \begin{pmatrix}
     T_{\myn, (\ell-1) m}^\dagger(\theta', \phi') \\
     S_{\myn, \ell m}^\dagger(\theta', \phi')
    \end{pmatrix} \,,
 }
where in order to simplify the subsequent expressions we defined the energy
 \es{Edef}{
  E_{\myn, \ell} &\equiv \sqrt{\ell^2 - \myn^2/{4}} \,.
 }
One can also perform the sum over $m$ using the explicit expressions \eqref{SpinorHarmonics} for the spinor harmonics.  To understand the result, it is simpler to first consider a similar sum over $m$ for the scalar monopole harmonics:
 \es{ScalarAddition}{
  \sum_{m = -\ell}^\ell Y_{\frac{\myn}{2}, \ell m}(\theta, \phi) Y^*_{\frac{\myn}{2}, \ell m}(\theta', \phi') 
   &= e^{-i \myn  \Theta} F_{\frac{\myn}{2}, \ell}(\gamma) \,, \\
    F_{\frac{\myn}{2}, \ell}(\cos \gamma) &\equiv 
      \sqrt{\frac{2 \ell + 1}{4 \pi}}  Y_{\frac{\myn}{2}, \ell, -\frac{\myn}{2}}(\gamma, 0) 
       = \frac{2 \ell+1}{4 \pi 2^{\myn/2}} ( 1 + \cos \gamma)^{\myn/2} P_{\ell - \frac{\myn}{2}}^{(0, \myn)}(\cos \gamma) \,,
 } 
where $\gamma$ is the angle between two points on $S^2$,
 \es{gammaDef}{
  \cos \gamma &= \hat x \cdot \hat x' = \cos \theta \cos \theta'  + \sin \theta \sin \theta' \cos (\phi - \phi') \,, 
 }
and the phase factor $e^{i \Theta}$ is 
 \es{ThetaDef}{
  e^{i \Theta} \cos \frac \gamma 2 &= 
   \cos \frac \theta 2 \cos \frac {\theta'}2 
    + e^{-i (\phi - \phi')} \sin \frac \theta 2 \sin \frac{\theta'}{2} \,.
 }
The addition formula \eqref{ScalarAddition} is the generalization of the notion of zonal spherical harmonics to the case of non-vanishing monopole flux.  Note that when $\myn>0$, the sum over $m$ yields an expression that, up to a phase, depends only on the relative angle between the two points on $S^2$.  Fortunately, this phase will not play a role in the computation below.

The addition formulas that generalize \eqref{ScalarAddition} to the case of the spinor harmonics are more complicated because of the spinor indices, and they will not be reproduced here.  They give
 \es{GqSimp}{
   G_n(r, r') &=  \sum_{\ell = \frac{\myn}{2}}^\infty \frac{i e^{-E_{\myn, \ell} \abs{\tau - \tau'}} e^{- i \myn \Theta} }{2}
  \Biggl[ 
   Q^1_{\myn, \ell} (\cos \gamma) \left(\hat x - \hat x' \right) \cdot \vec{\sigma} 
    + \sgn (\tau - \tau') \\ &\times \biggl(Q^2_{\myn, \ell} (\cos \gamma) {\bf 1} + Q^3_{\myn, \ell}  (\cos \gamma) \left(\hat x + \hat x' \right) \cdot \vec{\sigma}
     + i Q^4_{\myn, \ell} (\cos \gamma) (\hat x \times \hat x') \cdot \vec{\sigma}  \biggr)
  \Biggr]  \,,
 }
where the coefficients $Q^i_{\myn, \ell}(x)$ (where $x = \cos \gamma$) can all be expressed in terms of the function
 \es{Qdef}{
  Q_{\myn, \ell}(x) &\equiv \frac{1}{2 \ell + 1} F_{\myn, \ell}(x) - \frac{1}{2 \ell-1} F_{\myn, \ell-1} (x)
 }
as
 \es{GotQ4}{
  Q^1_{\myn, \ell}(x) &= \frac{E_{\myn, \ell}}{1 - x}  Q_{\myn, \ell} (x) \,, \qquad
    Q^3_{\myn, \ell}(x) =Q_{\myn, \ell}'(x)  \,, \\
  Q^2_{\myn, \ell}(x) &= \frac{\myn}{2} Q_{\myn, \ell} (x) \,, \qquad
    Q^4_{\myn, \ell}(x) = \frac{\myn}{2(1 + x)} Q_{\myn, \ell}(x) \,.
 }
Explicit care must be taken for $\ell = \myn/2$ where one should replace \eqref{Qdef} by 
 \es{QMin}{
  Q_{\myn, \myn/2}(x) = \frac{(1 + x)^{\myn/2}}{2^{2 + \frac{\myn}{2}} \pi} \,.
 }
In other words, only the first term in the expression for $Q_{\myn, \ell}(x)$ in \eqref{Qdef} should be considered in this case.  The reader is referred to Appendix~\ref{FERMIONCHECK} for a check that when $\myn=0$, the Green's function \eqref{GqSimp} agrees with what one would expect from a conformal transformation of the flat space Green's function.

Before moving on, let's discuss a few properties of the function $Q_{\myn, \ell}(x)$ that will be important later on.  In terms of Jacobi polynomials, the definition \eqref{Qdef} reads
 \es{QJacobi}{
  Q_{\myn, \ell}(x) = \frac{(1+x)^{\myn/2}}{2^{2+\frac{\myn}{2}} \pi} \left[P_{\ell-\frac{\myn}{2}}^{(0, \myn)} (x) - P_{\ell-1 - \frac{\myn}{2} }^{(0, \myn)}(x) \right] \,.
 }
Using an identity that relates the difference of two Jacobi polynomials to a third Jacobi polynomial of a different rank, $Q_{\myn, \ell}(x)$ can also be expressed as
 \es{QNewJacobi}{
  Q_{\myn, \ell}(x) =  -\frac{\ell (1+x)^{\myn/2} (1-x)}{2^{2+\frac{\myn}{2}} \pi (\ell-\myn/2)}  P_{\ell-1 - \frac{\myn}{2}}^{(1, \myn)}(x) \,.
 }
Quite remarkably, the Jacobi polynomial in \eqref{QNewJacobi} can be traded for a scalar monopole harmonic with $n-1$ units of gauge flux:
 \es{QMonopole}{
  Q_{\myn, \ell}(\cos \theta) = -\sqrt{\frac{\ell (1 - \cos \theta)}{4 \pi (\ell^2 - \myn^2/4)}}e^{i \phi} Y_{\frac{\myn-1}{2}, \ell-\frac 12, -\frac{\myn+1}{2}} (\theta, \phi) \,.
 }
Because of this relation, the function $Q_{\myn, \ell}(x)$ satisfies a second order differential equation that follows from the eigenvalue equation \eqref{LSqEvalue} for the monopole harmonics:
 \es{QDiffEq}{
  Q_{\myn, \ell}''(x) + \frac{1}{1+x} Q_{\myn, \ell}'(x) + \frac{1}{1 -x^2} \left[ \ell^2 - \frac{\myn^2}{2(1+x)} \right] Q_{\myn, \ell}(x)  = 0 \,.
 }
This equation also holds in the case $\ell = \myn/2$ where  one should use \eqref{QMin} instead of \eqref{Qdef}.  

Since the function $Q_{\myn, \ell}$ arose from describing fermions with total angular momentum $j = \ell - 1/2$, it may not be very surprising that it can be expressed in terms of a monopole harmonic with angular momentum equal to $j$.  It is less clear why the scalar monopole harmonic that appears in \eqref{QMonopole} experiences one fewer units of magnetic flux than the spinor harmonics.  In particular, when $\myn = 1$, $Q_{1, \ell}(x)$ can be expressed in terms of the usual spherical harmonics $Y_j^{-1}$, or equivalently in terms of the associated Legendre polynomials $P_\ell^1(x)$.

\section{Evaluating the gauge field kernel}
\label{KERNEL}

\subsection{General formulas}

Using \eqref{GotKq}--\eqref{NonZeroEvalues} and \eqref{Qdef}--\eqref{GotQ4}, as well as the explicit formulas for the spinor and vector harmonics, one finds
 \es{KqVV}{
  K^{\myn, E}_{\ell}(\omega)  &=  \frac{(4 \pi)^2}{(2 \ell+1) \ell (\ell+1)} \sum_{\ell', \ell''} \frac{E_{\myn, \ell'} +  E_{\myn, \ell''}}{\omega^2 + (E_{\myn, \ell'}+ E_{\myn, \ell''})^2} 
    \left[\frac { {\cal I}^E_1}2 +  E_{\myn, \ell'} E_{\myn, \ell''} {\cal I}^E_2 \right]\,, \\
  K^{\myn, B}_\ell (\omega) &= \frac{(4 \pi)^2}{(2 \ell+1) \ell (\ell+1)} \sum_{\ell', \ell''} \frac{E_{\myn, \ell'} +  E_{\myn, \ell''}}{\omega^2 + (E_{\myn, \ell'}+ E_{\myn, \ell''})^2}
     \left[  \frac { {\cal I}^B_1}2 + E_{\myn, \ell'} E_{\myn, \ell''} {\cal I}^B_2 \right] \,,
 }
with
 \es{IDefs}{
  {\cal I}^E_1 &= 2 \int_{-1}^1 dx \, \left[\frac{2 \ell (\ell+1) F_{0, \ell}+ (1-x) F_{0, \ell}' }{4(1+x)}
     \myn^2 Q_{\myn, \ell'} Q_{\myn, \ell''} -(1 - x^2) F_{0, \ell}'  Q_{\myn, \ell'}' Q_{\myn, \ell''}'   \right] \,, \\
  {\cal I}^E_2 &= -\int_{-1}^1 dx \, \frac{1+x}{1-x} F_{0, \ell}' Q_{\myn, \ell'} Q_{\myn, \ell''} \,, \\
  {\cal I}^B_1 &= 2 \int_{-1}^1 dx\, \left(F_{0, \ell}' - (1 - x) F_{0, \ell}'' \right) \left(\frac{\myn^2}{4} Q_{\myn, \ell'} Q_{\myn, \ell''} - (1+x)^2 Q_{\myn, \ell'}' Q_{\myn, \ell''}' \right) \,,  \\
  {\cal I}^B_2 &= \int_{-1}^1 dx\, \left(F_{0, \ell}' + (1 + x) F_{0, \ell}'' \right)  Q_{\myn, \ell'} Q_{\myn, \ell''} \,.
 }
Since both $F_{0, \ell}$ and $Q_{\myn, \ell}$ obey second order differential equations and since all the expressions in \eqref{IDefs} are symmetric under exchanging $\ell'$ and $\ell''$, \eqref{IDefs} can be brought into the canonical form
 \es{IGeneral}{
  \int_{-1}^1 dx\, \left[\alpha(x) F_{0, \ell}(x) + \beta(x) F_{0, \ell}'(x) \right] Q_{\myn, \ell'}(x) Q_{\myn, \ell''}(x) \,,
 } 
with some functions $\alpha(x)$ and $\beta(x)$ that depend on which expression in \eqref{IDefs} we are considering.  This canonical form is useful when evaluating the integrals with respect to $x$.

\subsection{Particular cases}

\subsubsection{$\myn=0$}

Let us start by evaluating the kernel $K^{0, E}_\ell(\omega)$.  By ``inspection'' of the integrals in \eqref{IDefs} for many values of $\ell$, $\ell'$, and $\ell''$, it is possible to guess the general formula
 \es{KE0}{
  K^{0, E}_\ell(\omega) = \frac{1}{2 \pi \ell (\ell+1)} \sum_{\ell', \ell''}  \frac{(\ell' + \ell'') ((\ell' - \ell'')^2 - \ell^2) 
    ((\ell' + \ell'')^2 - \ell (\ell+1)) (\ell + \ell' + \ell'' - 1)}{(\omega^2 + (\ell' + \ell'')^2) (\ell + \ell' + \ell'')} \\
   \times \begin{pmatrix}
    \ell'-1 & \ell''-1 & \ell-1 \\
    0 & 0 & 0
   \end{pmatrix}^2  \,,
 }
where factor on the second line is the square of a $3$-$j$ symbol. At fixed $\ell'$, the summation over $\ell''$ is of course convergent because the $3$-$j$ symbol vanishes when $\ell''>\ell' + \ell-1$.  However, the remaining summation over $\ell'$ is divergent because the terms in the sum approach the constant value $-1/(4 \pi)$ at large $\ell'$.  This divergence can be regularized by adding and subtracting $1/(4 \pi)$ from each term and then using the zeta-function identity $\sum_{\ell'=1}^\infty 1 = \zeta(-1) = -1/2$:
 \es{KE0Reg}{
  K^{0, E}_\ell(\omega) = \sum_{\ell' = 1}^\infty \left( a_{\ell, \ell'}(\omega) + \frac{1}{4 \pi} \right)
   - \sum_{\ell'=1}^\infty \frac{1}{4 \pi}  = \frac{1}{8 \pi} +  \sum_{\ell' = 1}^\infty \left( a_{\ell, \ell'}(\omega) + \frac{1}{4 \pi} \right)  \,.
 } 
The summation over $\ell'$ is now convergent and can be performed explicitly at fixed $\ell$.  For instance, for $\ell=1$ we have
 \es{K0E1}{
  K^{0,E}_1(\omega) &=  \frac{1}{8 \pi} + \sum_{\ell'=1}^\infty \frac{4 \ell'^2 (1 + \omega^2) - \omega^2}
    {4 \pi (4 \ell'^2 - 1) (4 \ell'^2 + \omega^2)} =(2 + \omega^2)  \frac{\omega \coth \frac{\pi \omega}{2}  }
    {16 (1 + \omega^2)} \,.
 }
Similarly,
 \es{MoreKEs}{
  K^{0, E}_2(\omega) &= (6 + \omega^2) \frac{(1 + \omega^2) \tanh \frac{\pi \omega}{2}}
    {16 \omega (4 + \omega^2)} \,, \\
  K^{0, E}_3(\omega) &= (12 + \omega^2) \frac{\omega(4 + \omega^2) \coth \frac{\pi \omega}{2}}
    {16 (1 + \omega^2) (9 + \omega^2)}  \,,
 } 
and so on.  It is not hard to see that in general\footnote{The same expressions for $K^{0, E}_\ell(\omega)$ and $K^{0, B}_\ell(\omega)$ were obtained in collaboration with Subir Sachdev as part of a similar computation in the $\CP^N$ model.}
 \es{KEFinal}{
  K^{0, E}_\ell(\omega) &= \frac {\ell(\ell+1) + \omega^2 }2  D^0_\ell(\omega) \,,
 } 
where 
 \es{D0Def}{
  D_\ell^0(\omega) = \abs{ \frac{\Gamma\left((\ell + 1 + i \omega) / 2 \right)}{4\Gamma\left((\ell + 2 + i \omega) / 2 \right)}  }^2
 }
is the scalar field kernel appearing in eq.~(35) of \cite{Pufu:2013eda}.

One can perform a similar analysis for $K^{0, B}(\omega)$.  By inspection of \eqref{KqVV}--\eqref{IDefs}, I found
 \es{KB0}{
  K^{0, B}_\ell(\omega) = \frac{1}{2 \pi \ell (\ell+1)} \sum_{\ell', \ell''} \frac{(\ell' - \ell'')^2 (\ell' + \ell'') (\ell + 1 - \ell' - \ell'') (\ell + \ell' + \ell'')}
    {\omega^2 + (\ell' + \ell'')^2} \\
   \times \begin{pmatrix}
    \ell'-1 & \ell''-1 & \ell \\
    0 & 0 & 0
   \end{pmatrix}^2 \,.
 }
The sums over $\ell'$ are again divergent but can be regularized precisely as in \eqref{KE0Reg}.  A few particular cases give
 \es{KB0Final}{
  K^{0, B}_1(\omega) &= \frac{(1 + \omega^2)\tanh \frac{\pi \omega}{2}}{16 \omega} \,, \\
  K^{0, B}_2(\omega) &= \frac{\omega (4 + \omega^2) \coth \frac{\pi \omega}{2}}{16 (1 + \omega^2)} \,,\\
  K^{0, B}_3(\omega) &= \frac{(1 + \omega^2) (9 + \omega^2) \tanh \frac{\pi \omega}{2}}{16 \omega(4 + \omega^2)} \,,
 } 
and so on.  From these expressions one can guess the general formula
 \es{KBFinal}{
  K^{0, B}_\ell(\omega) = \frac{ \ell^2 + \omega^2}{2} D^0_{\ell-1} (\omega) \,.
 }
The quantity that appears in \eqref{FCorrected} is the product $K^{0, E}_\ell(\omega) K^{0, B}_\ell(\omega)$, which is given by
 \es{KProduct}{
  K^{0, E}_\ell(\omega) K^{0, B}_\ell(\omega) = \frac{\ell(\ell+1) + \omega^2}{256} \,.
 }

\subsubsection{$\myn=1$}

When $\myn = 1$, the expressions \eqref{IDefs} written in the canonical form \eqref{IGeneral} are
 \es{IHalf}{
  {\cal I}^E_1 &= \left(J_1 - J_2 \right) \left[\ell (\ell+1) - \ell'^2 -\ell''^2 + \frac 12  \right]\,, \\
  {\cal I}^E_2 &= -  J_1 - J_2\,, \\
  {\cal I}^B_1 &=  \ell(\ell+1)  \left[  \ell(\ell+1)  J_0 - 2J_2 \right]
    -  \left[ J_1 - J_2 + \ell(\ell+1) J_0 \right] \left[ \ell'^2 + \ell''^2 - \frac 12  \right]  \,, \\
  {\cal I}^B_2 &=  J_1 + J_2 - \ell(\ell+1)  J_0   \,,
 }
where $J_i$ are integrals involving $F_{0, \ell}$, $Q_{1, \ell'}(x)$, and $Q_{1, \ell''}(x)$:
 \es{JDefs}{
  J_0(\ell, \ell', \ell'') &= \int_{-1}^1 dx\, \frac{1}{1-x} F_{0, \ell}(x) Q_{1, \ell'}(x) Q_{1, \ell''}(x) \,, \\
  J_1(\ell, \ell', \ell'') &= \int_{-1}^1 dx\, \frac{1}{1-x} F_{0, \ell}'(x) Q_{1, \ell'}(x) Q_{1, \ell''}(x) \,, \\
  J_2(\ell, \ell', \ell'') &= \int_{-1}^1 dx\, \frac{x}{1-x} F_{0, \ell}'(x) Q_{1, \ell'}(x) Q_{1, \ell''}(x) \,.
 }
One needs to evaluate these integrals.  In doing so it is helpful to recall that \eqref{QMonopole} implies that $Q_{1, \ell}(x)$ is expressible in terms of the associated Legendre polynomials:
 \es{QLegendre}{
  Q_{1, \ell}(x) = \frac{\sqrt{2} \ell}{\pi (2 \ell-1)(2\ell+1)}  \sqrt{1-x}  P_{\ell-1/2}^1(x) \,.
 }
We should also recall that $F_{0, \ell}$ is proportional to a Legendre polynomial:
 \es{F0P}{
  F_{0, \ell}(x) = \frac{2 \ell+1}{4 \pi} P_\ell(x) \,.
 }
When substituting \eqref{F0P} into $J_1$ and $J_2$ we should also make use of the following formulas relating the derivative of the Legendre polynomials to associated Legendre polynomials:
 \es{Pder}{
  P_{\ell}'(x) &= \frac 12 \left[P_{\ell+1}^2(x) + \ell(\ell+1) P_{\ell+1}^0(x) \right] \,, \\
  x P_{\ell}'(x) &= \frac 12 \left[P_{\ell}^2(x) + \ell(\ell+1) P_{\ell}^0(x) \right] \,.
 }
The $J_i$ then become integrals over products of three associated Legendre polynomials;  they can be evaluated using the formula
 \es{ThreeJFormula}{
  \frac 12 \int_{-1}^1 dx\, P_{\ell_1}^{m_1}(x) P_{\ell_2}^{m_2}(x) P_{\ell_3}^{m_3}(x) 
   = (-1)^{m_3} 
    \begin{pmatrix}
     \ell_1 & \ell_2 & \ell_3 \\
     0 & 0 & 0
    \end{pmatrix}
     \begin{pmatrix}
     \ell_1 & \ell_2 & \ell_3 \\
     m_1 & m_2 & -m_3
    \end{pmatrix}
    \prod_{i=1}^3 \sqrt{ \frac{(\ell_i +m_i)!}{(\ell_i - m_i)!}} \,,
 }
which holds for $m_3 = m_1 + m_2$.   The resulting expressions for $J_+$, $J_-$, and $J_0$ are relatively messy, but each of these quantities can be expressed in terms of $3$-$j$ symbols that can be easily evaluated using a computer program:
 \es{GotJ0}{
   J_0 \left(\ell, \frac 12 + \ell_1, \frac 12 + \ell_2\right) = 
   -\frac{\left(\ell_1 +\frac 12\right) \left(\ell_2 +\frac 12\right) (2\ell+1)}{16 \pi^3 \sqrt{\ell_1(\ell_1+1)  \ell_2(\ell_2+1)}}
    \begin{pmatrix}
     \ell & \ell_1 & \ell_2 \\
     0 & 0 & 0
    \end{pmatrix}
     \begin{pmatrix}
     \ell & \ell_1 & \ell_2 \\
     0 & 1 & -1 
    \end{pmatrix} \,,
 }
 \es{GotJ1}{
 J_1 \left(\ell, \frac 12 + \ell_1, \frac 12 + \ell_2\right) &=   \frac{\left(\ell_1 +\frac 12\right) \left(\ell_2 +\frac 12\right) (2\ell+1) \sqrt{\ell(\ell+ 1)}}{32 \pi^3 \sqrt{\ell_1(\ell_1+1)  \ell_2(\ell_2+1)}}
    \begin{pmatrix}
     \ell+1 & \ell_1 & \ell_2 \\
     0 & 0 & 0
    \end{pmatrix}  \\
    &\times 
     \left[ \sqrt{(\ell+2)(\ell+3)}  \begin{pmatrix}
      \ell+1 & \ell_1 & \ell_2 \\
     -2 & 1 & 1
    \end{pmatrix} - \sqrt{\ell(\ell+ 1)}
      \begin{pmatrix}
      \ell+1 & \ell_1 & \ell_2 \\
     0 & 1 & -1
    \end{pmatrix}  \right]  \,, \\
 } 
  \es{GotJ2}{
 J_2 \left(\ell, \frac 12 + \ell_1, \frac 12 + \ell_2\right) &=   \frac{\left(\ell_1 +\frac 12\right) \left(\ell_2 +\frac 12\right) (2\ell+1) \sqrt{\ell(\ell+ 1)}}{32 \pi^3 \sqrt{\ell_1(\ell_1+1)  \ell_2(\ell_2+1)}}
    \begin{pmatrix}
     \ell & \ell_1 & \ell_2 \\
     0 & 0 & 0
    \end{pmatrix}  \\
    &\times 
     \left[ \sqrt{(\ell-1)(\ell+2)}  \begin{pmatrix}
      \ell & \ell_1 & \ell_2 \\
     -2 & 1 & 1
    \end{pmatrix} - \sqrt{\ell(\ell+ 1)}
      \begin{pmatrix}
      \ell & \ell_1 & \ell_2 \\
     0 & 1 & -1
    \end{pmatrix}  \right]  \,.
 }
With \eqref{GotJ0}--\eqref{GotJ2} and \eqref{IHalf}, one can now evaluate $K^{1, E}_\ell(\omega)$ and $K^{1, B}_\ell(\omega)$ using \eqref{KqVV}.  I checked that plugging \eqref{GotJ0}--\eqref{GotJ2} into \eqref{IHalf} yields the same answers as performing the integrals in \eqref{IDefs} explicitly.

In \eqref{KqVV}, the sums over $\ell'$ and $\ell''$ run from $1/2$ to infinity.   While there are no problems with using \eqref{IHalf} when $\ell', \ell'' > 1/2$, extra care must be taken when $\ell' = 1/2$ or $\ell'' = 1/2$.  It is not hard to see using \eqref{QMin} that the contribution to $K_\ell^{1, E}(\omega)$ from $\ell' = 1/2$ and/or $\ell'' = 1/2$ vanishes, while the contribution to $K_\ell^{1, B}(\omega)$ is given by
 \es{KlBExtra}{
  -\frac 1{4 \pi} \frac{\sqrt{\ell(\ell+1)}}{\omega^2 + \ell(\ell+1)} \,.
 }
So one can restrict the sums in \eqref{KqVV} to run over $\ell', \ell'' > 1/2$ and add \eqref{KlBExtra} to $K_\ell^{1, B}(\omega)$.

\subsection{Numerics for $\myn = 1$}

\begin {figure} [htb]
  \center\includegraphics [width=0.6\textwidth] {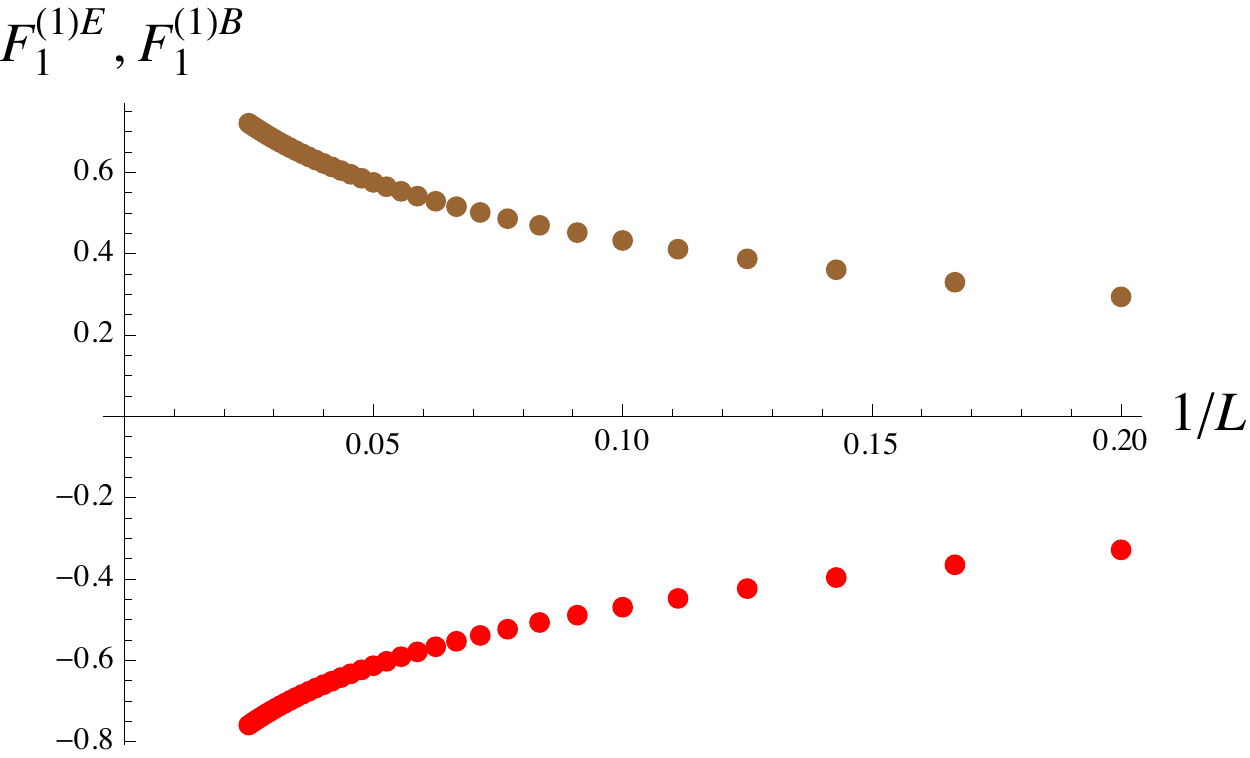}
  \caption {The quantities $F^{(1) E}_{1}$ (brown) and $F^{(1) B}_{1}$ (red) evaluated with the relativistic cutoff \eqref{IsotropicCutoff}.  They each diverge logarithmically as $L \to \infty$.  \label {FEBPlot}}
\end {figure}

\begin {figure} [htb]
  \center\includegraphics [width=0.6\textwidth] {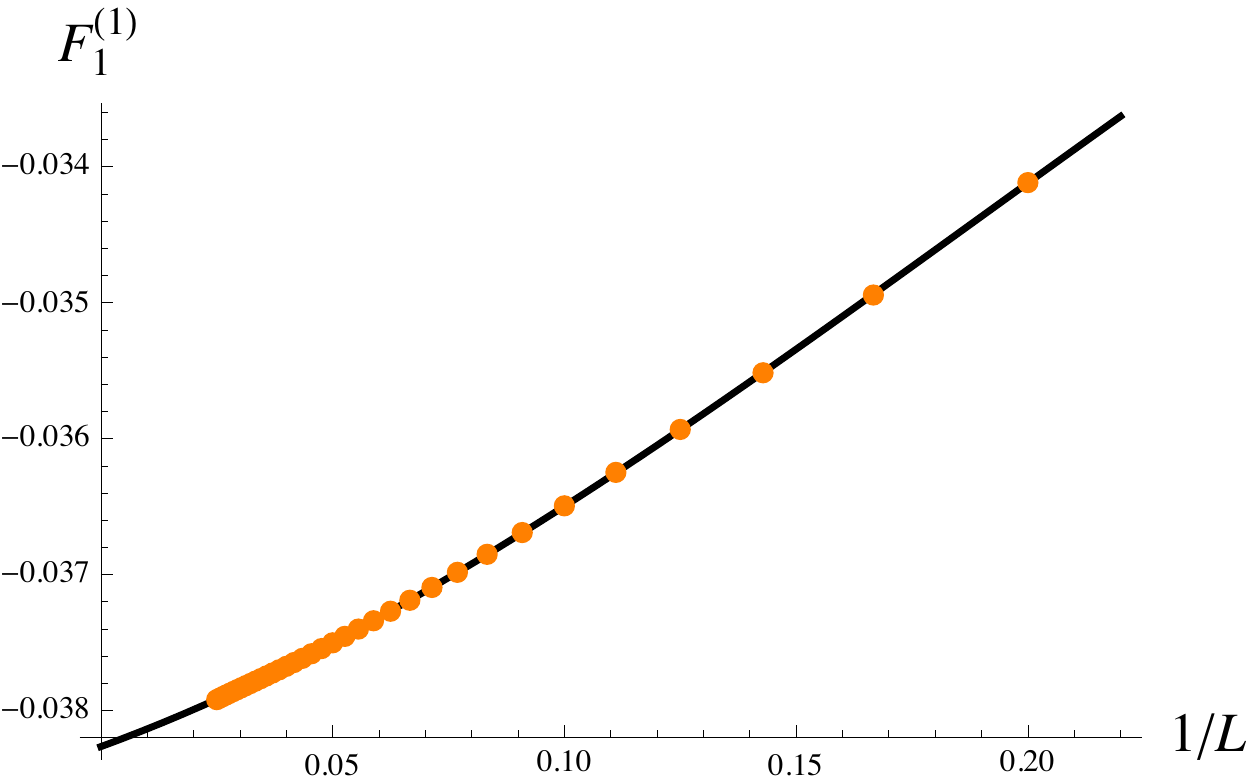}
  \caption {The correction $ F^{(1)}_1 =  F^{(1)E}_{1} +  F^{(1)B}_{1}$ to the free energy evaluated by summing up the expressions in \eqref{deltaFEB} with the relativistic cutoff \eqref{IsotropicCutoff}. The orange points are obtained by evaluating \eqref{deltaFEB} numerically, and the solid line is a cubic fit.  \label {FPlot}}
\end {figure}

One can calculate the corresponding contributions to the free energy \eqref{FCorrected} when $\myn = 1$:
 \es{deltaFEB}{
   F^{(1)E}_{1} = \frac 12 \int \frac{d\omega}{2 \pi} \sum_{\ell=1}^\infty (2 \ell+1) \log \frac{K_\ell^{1, E}(\omega) }{K_\ell^{0, E}(\omega)} \,, \\
   F^{(1)B}_{1} = \frac 12 \int \frac{d\omega}{2 \pi} \sum_{\ell=1}^\infty (2 \ell+1) \log \frac{K_\ell^{1, B}(\omega) }{K_\ell^{0, B}(\omega)} \,.
 }
These expressions are defined so that the total $O(N_f^0)$ correction to the free energy is $F^{(1)}_1 = F^{(1)E}_{1} + F^{(1)B}_{1}$.  Imposing a relativistic cutoff on the summations and integrations in \eqref{deltaFEB} by restricting $\ell$ and $\omega$ in the range 
 \es{IsotropicCutoff}{
   \ell(\ell+1) + \omega^2 \leq L(L+1) \,,
  } 
for some cutoff energy scale $L$, I find that both $F^{(1)E}_{1}$ and $ F^{(1) B}_{1}$ diverge logarithmically with $L$: see Figure~\ref{FEBPlot}.
However, as can be seen in Figure~\ref{FPlot}, the logarithmic divergence cancels out from the sum $F^{(1)}_1$.

Extrapolating to $L = \infty$, I get
 \es{F1}{
   F_1^{(1)} \approx -0.0383 \,.
 }
This is the sought-after correction in \eqref{Expansion} to the ground state energy on $S^2$ in the presence of one unit of magnetic flux through the two-sphere.

\section{Discussion}
\label{DISCUSSION}

The main result of this paper is that the scaling dimension of the monopole operator that inserts one unit of magnetic flux in QED$_3$ with $N_f$ fermion flavors is
 \es{FinalResult}{
  0.265 N_f - 0.0383 + O(1/N_f) \,,
 }
which is obtained by combining the leading order result \cite{Borokhov:2002ib} given in Table~\ref{F0Table} with \eqref{F1}.  The $O(1)$ correction in \eqref{FinalResult} was obtained by performing a Gaussian integral over the fluctuations of the gauge field on $S^2 \times \R$ around a background constant magnetic flux of $2 \pi$ that is uniformly distributed throughout the $S^2$.  

A striking feature of this computation is that the fermion Green's function on $S^2 \times \R$ in the presence of $\myn$ units of magnetic flux (as well as the kernel that appears in the quadratic action for the gauge field fluctuations, which can be written in terms of the fermion Green's function) can be expressed in terms of the {\em spinless} $S^2$ monopole spherical harmonics of \cite{Wu:1976ge,Wu:1977qk} in the presence of $n-1$ units of magnetic flux.  This major simplification is what made the computation in this paper at all possible, and it would be interesting to understand its origin from a more conceptual point of view.  In the case $n=1$ analyzed here, the fermion Green's function is particularly simple because it is expressible in terms of the usual spherical harmonics on $S^2$, or equivalently in terms of certain associated Legendre polynomials.

Another feature of the computation presented in this paper is an exact cancellation of UV divergences.  Because of symmetry under rotations on $S^2$ and translations along $\R$, the quadratic action for the gauge field fluctuations is diagonalized by a combination of vector spherical harmonics on $S^2$ and plane waves in the $\R$ direction.  There are two gauge-invariant sectors depending on the properties of the vector spherical harmonics on $S^2$: one sector involving vector harmonics with zero curl (the ``$E$-modes''), and one sector involving vector harmonics with zero divergence (the ``$B$-modes'').  The contribution of each of these two sectors to the $S^2$ ground state energy can be evaluated independently.  Strikingly, each contribution is logarithmically UV divergent, but the divergences cancel exactly when the $E$-modes and the $B$-modes are added up together.  This cancellation is far from obvious at intermediate stages of the computation, and it therefore provides a check on the method used here.

A few comments on the significance of the result \eqref{FinalResult} are in order.  It can be noticed that the $O(1)$ term in \eqref{FinalResult} is small compared to the coefficient of the leading term proportional to $N_f$.  Since the expansion \eqref{FinalResult} is likely to be an asymptotic series, the smallness of the $O(1)$ term suggests that the approximation \eqref{FinalResult} might be accurate even for small $N_f$.  It is then reasonable to use this approximation to extract estimates for the upper limits on the number of fermion flavors below which compact QED$_3$ confines and below which the naive CFT limit $e^2 \to \infty$ in non-compact QED$_3$ might break down.

As mentioned in the introduction, in compact QED$_3$ monopole operators can proliferate and lead to confinement provided that they are relevant in the RG sense, i.e.~if their scaling dimensions are smaller than $3$.  From \eqref{FinalResult} it is easy to see that the $n=1$ monopole operator is relevant if $N_f \lsim 11.47$ and irrelevant otherwise.  I would therefore expect that compact QED$_3$ should be in a deconfined phase precisely if $N_f \geq 12$. This result is consistent with the lattice analysis of \cite{Armour:2011zx}, which shows that the monopoles proliferate for $N_f = 2$ and $N_f = 4$, as well with the results of \cite{Grover:2012sp}, where it is argued based on the $F$-theorem \cite{Myers:2010tj, Jafferis:2011zi, Klebanov:2011gs, Casini:2012ei} and the work of Vafa and Witten \cite{Vafa:1983tf,Vafa:1984xh} that confinement would not be possible for $N_f > 12$.

Lastly, one should stress that the computation presented in this paper was performed under the assumption that QED$_3$ flows to an infrared conformal fixed point obtained by taking $e^2 \to \infty$ in the action \eqref{ActionEuc}.  This  assumption is certainly correct at large $N_f$, where the whole RG flow can be studied perturbatively.  As discussed in the introduction, below a critical value of fermion flavors, $N_f \leq N_f^\text{crit}$, one expects the infrared physics to be significantly different from what the CFT limit $e^2 \to \infty$ would predict, the scenario supported by lattice data being that of spontaneous chiral symmetry breaking.  For $N_f \leq N_f^\text{crit}$ it is reasonable to speculate that something would go wrong with taking the limit $e^2 \to \infty$ in \eqref{ActionEuc}.   Prima facie evidence that this naive CFT limit is no longer appropriate would be that it leads to operator dimensions that violate the unitarity bound.  Such a situation is not uncommon.  In supersymmetric gauge theories, it is well-known that as one decreases the number of flavors it may happen that at some point certain operators hit the unitarity bound;  to continue decreasing the number of flavors, one must pass to a dual description of the CFT because the original one is no longer valid.   In non-supersymmetric QED$_3$ such a dual description may not be readily available, and for fewer fermion flavors new physics would be expected.

One can therefore estimate $N_f^\text{crit}$ by assuming that for $N_f \leq N_f^\text{crit}$ the naive CFT limit predicts unitarity bound violations.  The unit strength monopole operator is a Lorentz scalar, so in a unitary theory its scaling dimension should be no smaller than $1/2$.  Using \eqref{FinalResult}, one finds a unitarity bound violation for  $N_f \leq  2.03$.  Of the local operators that are polynomials in the fundamental fields, the one with lowest scaling dimension is $\psi_a^\dagger \psi_a$, for which \cite{Rantner:2002zz}
 \es{psidagpsi}{
   [\psi_a^\dagger \psi_a] = 2 - \frac{64}{3 \pi^2 N_f} + O(1/N_f^2) \,.
 } 
The scaling dimension of this operator violates the unitarity bound if $N_f \leq 1.44$.  While one should treat these bounds very cautiously, they suggest that the monopole operators are more constraining.  That the monopole operators are more important at small $N_f$ is consistent with the similar findings of \cite{Safdi:2012re, Yaakov:2013fza} in supersymmetric theories.  Lastly, it is intriguing that the estimate for $N_f^\text{crit}$ obtained this way is consistent with the lattice results, which suggest there is chiral symmetry breaking at $N_f = 2$, but not for larger values of $N_f$.

\section*{Acknowledgments}

I am extremely grateful to Subir Sachdev for collaboration on related projects and for many insightful conversations.  I also thank Liang Fu, Patrick Lee, Stefan Meinel, and Senthil Todadri for helpful discussions, as well as Ethan Dyer, Igor Klebanov, and Mark Mezei for their interest in my work and for their comments on preliminary versions of this paper.  This work was supported in part by a Pappalardo Fellowship in Physics at MIT and in part by the U.S. Department of Energy under cooperative research agreement Contract Number DE-FG02-05ER41360\@.

\appendix

\section{Monopole harmonics via conformal transformation}
\label{HARMONICS}

\subsection{Scalar harmonics}

The (scalar) monopole spherical harmonics $Y_{q, \ell m}(\theta, \phi)$ of \cite{Wu:1977qk,Wu:1976ge} can be used as a basis for the angular dependence of a charged scalar field in a background with $n = 2q$ units of monopole flux sitting at the origin of $\R^3$.  Just like the usual spherical harmonics, they are eigenfunctions of the angular momentum operators $\vec{L}^2$ and $L_3$ with eigenvalues given by
 \es{LLzEvalues}{
  {\vec L}^2 Y_{q, \ell m} = \ell(\ell + 1) Y_{q, \ell m} \,, \qquad L_3 Y_{q, \ell m} = m Y_{q, \ell m} \,.
 }
If the electromagnetic potential is given by \eqref{BackgroundGauge}, the angular momentum operators can be written as
 \es{AngMom}{
  L_3 &= - i \partial_\phi - q \,, \\
  {\vec L}^2 &= -{\nabla}^2 + \frac{2 q}{\sin^2 \theta} (\cos \theta - 1) L_3 \,,
 }
where $\nabla^2$ is the usual Laplacian on the two-sphere.  It will be useful for us to write 
 \es{SeparationVariables}{
  Y_{q,\ell m}(\theta, \phi) = \Theta_{q, \ell m} (\cos \theta) e^{i (m + q) \phi} \,.
 }
The ${\vec L}^2$ eigenvalue equation can then be written as
 \es{LSqEvalue}{
  - \partial_x \left[(1 - x^2) \partial_x \Theta_{q, \ell m}(x) \right]
   + \frac{m^2 + q^2 + 2 q m x}{1 - x^2} \Theta_{q, \ell m}(x) = \ell(\ell+1) \Theta_{q, \ell m}(x) \,.
 }
The regular solution of this equation can be given in terms of the Jacobi polynomials: 
 \es{ThetaJacobi}{
  \Theta_{q, \ell m} (x) = 2^{m-1} \sqrt{\frac{(2 \ell + 1) (\ell - m)! (\ell + m)!}{\pi (\ell - q)! (\ell + q)!}} \sqrt{\frac{(1 + x)^{q-m}}{ (1-x)^{q+m}}}
   P_{\ell + m}^{(-q-m, q-m)} (x) \,.
 }
The normalization factor in \eqref{ThetaJacobi} is chosen such that $\abs{Y_{q, \ell m}(\theta, \phi)}^2$ integrates to unity over the two-sphere.

Let's denote the scalar monopole harmonics $Y_{q, \ell m}$ as $| \ell, m \rangle$, suppressing for now the extra label $q$.  They transform in the spin-$\ell$ representation of the $SO(3)$ rotation group.  In these conventions, $q$ is allowed to take half-integer values, and $\ell - \abs{q} \geq 0$ is an integer.

\subsection{Spinor and vector harmonics}

A basis of functions for a charged field of spin $s$ in the same monopole background can be constructed using the angular momentum addition rules.  The $(2s+1)$ components of this field can be written as $|s, m_s \rangle$, with $-s \leq m_s \leq s$, and they transform in the spin-$s$ representation of $SO(3)$.  A basis of monopole spinor harmonics is given by
 \es{AngMomAddition}{
  | \ell, s, j, m_j \rangle =(-1)^{-\ell + s - m_j} \sqrt{2 j + 1} \sum_{m_s = -s}^s   
  \begin{pmatrix} 
    \ell & s & j \\ 
    m_j - m_s & m_s & -m_j
   \end{pmatrix}
    | \ell, m_j - m_s \rangle \otimes | s, m_s \rangle \,,
 }
where I wrote down the Clebsch-Gordan coefficients explicitly in terms of the 3-$j$ symbols.

In particular, for $s = 1/2$ there are two sets of modes
 \es{SpinorHarmonics}{
    T_{q, \ell m}(\theta, \phi) &=  \left| \ell, \frac 12 , \ell + \frac 12 , m + \frac 12 \right\rangle = 
   \begin{pmatrix}
      \sqrt{\frac{\ell + m + 1}{2 \ell + 1}} Y_{q, \ell m} (\theta, \phi) \\
    \sqrt{\frac{\ell - m}{2 \ell + 1}} Y_{q, \ell (m+1)}(\theta, \phi) 
   \end{pmatrix}  \,, \\
  S_{q, \ell m}(\theta, \phi) &=  \left| \ell, \frac 12 , \ell - \frac 12 , m + \frac 12 \right\rangle = 
   \begin{pmatrix}
    - \sqrt{\frac{\ell - m}{2 \ell + 1}} Y_{q, \ell m}(\theta, \phi)  \\
    \sqrt{\frac{\ell + m + 1}{2 \ell + 1}} Y_{q, \ell (m+1)} (\theta, \phi) 
   \end{pmatrix} \,.
 }
(This is what's called $\phi_{j\pm 1/2, j m}$ in \cite{Borokhov:2002ib}.)  Note that for $T_{q, \ell m}$ we have $-\ell-1 \leq m \leq \ell$, and for $S_{q, \ell m}$ we have $-\ell \leq m \leq \ell -1$, as follows from the fact that in both cases $-j \leq m_j \leq j$.  If the monopole harmonics $Y_{q, \ell m}$ are normalized such that they have unit norm, then $T_{q, \ell m}$ and $S_{q, \ell m}$ will also have unit norm.

Similarly, when $q=0$, one can define the vector harmonics corresponding to $s = 1$:
 \es{VectorHarmonics}{
    U_{\ell m}(\theta, \phi) &=  \left| \ell, 1 , \ell - 1 , m  \right\rangle  \,, \\
    V_{\ell m}(\theta, \phi) &=  \left| \ell, 1 , \ell  , m \right\rangle \,, \\
    W_{\ell m}(\theta, \phi) &=  \left| \ell, 1 , \ell + 1 , m\right\rangle \,.
 }
In thinking about vector harmonics on $S^2$ it is customary to define
 \es{VectorS2}{
  \mathcal{X}_{i,\ell m}(\theta, \phi) &= \frac{1}{\sqrt{\ell (\ell + 1)}} \partial_i Y_{\ell m}(\theta, \phi) \,, \\
  \mathcal{Y}^i_{\ell m}(\theta, \phi) &= \frac{1}{\sqrt{\ell (\ell + 1)}} \frac{\epsilon^{ij}}{\sqrt{g}} \partial_j Y_{\ell m}(\theta, \phi)  \,,
 }
where the indices $i,j=\theta,\phi$, and $\epsilon^{\theta\phi} = 1$ is the unit antisymmetric tensor.   A straightforward calculation shows
 \es{UVWExplicit}{
  U_{(\ell+1) m}(\theta, \phi) &= 
    -\sqrt{\frac{\ell+1}{2 \ell +1}} Y_{\ell m}(\theta, \phi) d\tau + \sqrt{\frac{\ell}{2 \ell + 1}} {\cal X}_{\ell m}(\theta, \phi) \,, \\
  V_{\ell m}(\theta, \phi) &=  i {\cal Y}_{\ell m}(\theta, \phi) \,, \\
  W_{(\ell - 1) m} (\theta, \phi) &=\sqrt{\frac{\ell}{2 \ell + 1}} Y_{\ell m}(\theta, \phi) d \tau 
    + \sqrt{\frac{\ell+1}{2 \ell + 1}} {\cal X}_{\ell m}(\theta, \phi) \,.
 }

\section{A check:  Fermion Green's function at $\myn=0$}
 \label{FERMIONCHECK}

As a check, when $\myn=0$, the Green's function should be
 \es{GreenZeroExpect}{
  G_0(r, r') = \frac{i \abs{\vec{x}} \abs{\vec{x}'}}{4 \pi} \frac{\vec{\sigma} \cdot (\vec{x} - \vec{x}' )}{\abs{\vec{x} - \vec{x}'}^3} \,,
 }
where $\vec{x}$ was defined in \eqref{Frame} in terms of the coordinates on $S^2 \times \R$.  Apart from the $\abs{\vec{x}} \abs{\vec{x}'}$ factor, this expression is just the fermion Green's function in flat space, and $\abs{\vec{x}} \abs{\vec{x}'}$ is the conformal factor needed to map the theory from $\R^3$ to $S^2 \times \R$.  We can rewrite $G_0$ as
 \es{GreenZeroExpect2}{
  G_0(r, r') 
  = \frac{i}{4 \pi} \frac{\vec{\sigma} \cdot \left( e^{(\tau - \tau')/2} \hat x - e^{(\tau' - \tau)/2} \hat x' \right)}{\left( 2 \cosh (\tau - \tau') - 2 \cos \gamma \right)^{3/2}} \,.
 }

Let's now try to reproduce this expression from the spectral decomposition used in Section~\ref{GREEN}.  When $\myn=0$, \eqref{SpectralAgain} becomes
 \es{SpectralAgain0}{
  G_0(r, r') &=  i \sum_{\ell = 1}^\infty \sum_{m = -\ell}^{\ell-1} e^{-\ell \abs{\tau - \tau'}}
   \biggl[T_{0, (\ell-1) m}(\theta, \phi)  S_{0, \ell m}^\dagger(\theta', \phi') \theta(\tau' - \tau) \\
    &\qquad\qquad\qquad- S_{0, \ell m}(\theta, \phi) T_{0, (\ell-1) m}^\dagger(\theta', \phi')  \theta (\tau - \tau') \biggr] \,.
 }
The spinor addition are
 \es{SpinorAdditionZero}{
  \sum_{m = -\ell}^{\ell-1} T_{0, (\ell-1) m}(\theta, \phi)  S_{0, \ell m}^\dagger(\theta', \phi')
   &= \frac{\vec{\sigma}}{4 \pi}\cdot \left[  -\hat x' P_{\ell}'(\cos \gamma)
    + \hat x P_{\ell-1}'(\cos \gamma) \right] \,, \\
    \sum_{m = -\ell}^{\ell-1} S_{0, \ell m}(\theta, \phi) T_{0, (\ell-1) m}^\dagger(\theta', \phi')
   &= \frac{\vec{\sigma}}{4 \pi}\cdot \left[ -\hat x P_{\ell}'(\cos \gamma)
    + \hat x'  P_{\ell-1}'(\cos \gamma) \right] \,.
 }
Then
 \es{SpectralLegendre}{
  G_0(r, r') &=  \sum_{\ell = 1}^\infty \frac{i e^{-\ell \abs{\tau - \tau'}}}{4 \pi}  \vec{\sigma}\cdot
   \biggl[\left[  -\hat x' P_{\ell}'(\cos \gamma)
    + \hat x P_{\ell-1}'(\cos \gamma) \right]  (- \theta (\tau' - \tau) ) \\
    &\qquad\qquad\qquad+ \left[ -\hat x P_{\ell}'(\cos \gamma)
    + \hat x'  P_{\ell-1}'(\cos \gamma) \right] \theta (\tau - \tau') \biggr] \,.
 }

Next we should use the generating function for the Legendre polynomials:
 \es{GeneratingFn}{
  \frac{1}{\sqrt{1 - 2 t x + t^2}} = \sum_{n=0}^\infty P_n(x) t^n \,.
 }
Differentiating with respect to $x$ and setting $x = \cos \gamma$ and $t = e^{- \abs{\tau - \tau'}}$, one obtains
 \es{GeneratingFnDer}{
  \frac{e^{\abs{\tau - \tau'}/2}}{\left( 2 \cosh (\tau - \tau') - 2 \cos \gamma \right)^{3/2}} = \sum_{\ell=0}^\infty P_\ell'(\cos \gamma) e^{-\ell \abs{\tau - \tau'}} \,.
 }
Using this formula in \eqref{SpectralLegendre}, one further obtains
 \es{SpectralLegendre2}{
  G_0(r, r') &=  \frac{i}{4 \pi \left( 2 \cosh (\tau - \tau') - 2 \cos \gamma \right)^{3/2}}  \vec{\sigma}\cdot
   \biggl[\left(  -e^{\abs{\tau - \tau'}/2} \hat x' 
    +e^{-\abs{\tau - \tau'}/2} \hat x \right)  \theta(\tau' - \tau)  \\
    &\qquad\qquad\qquad- \left(-e^{\abs{\tau - \tau'}/2} \hat x 
    + e^{-\abs{\tau - \tau'}/2}\hat x'   \right) \theta(\tau - \tau') \biggr] \,.
 }
Combining the terms in the square brackets, the final result is
  \es{GotGreenZero}{
  G_0(r, r') 
  = \frac{i}{4 \pi} \frac{\vec{\sigma} \cdot \left( e^{(\tau - \tau')/2} \hat x - e^{(\tau' - \tau)/2} \hat x' \right)}{\left( 2 \cosh (\tau - \tau') - 2 \cos \gamma \right)^{3/2}} \,,
 }
which agrees with \eqref{GreenZeroExpect2}.  

\newpage

\bibliographystyle{ssg}
\bibliography{monopole}

\end{document}